\documentclass[12pt,preprint]{aastex}

\shorttitle{Active Galactic Nuclei in Massive Galaxies at High Redshift}
\shortauthors{Yamada et al.}

\newcommand{\gtsim}{\,\mbox{\raisebox{0.3ex}{$>$}\hspace{-0.8em}\raisebox{-0.7ex}{$\sim$}}\,}   
\newcommand{\ltsim}{\,\mbox{\raisebox{0.3ex}{$<$}\hspace{-0.8em}\raisebox{-0.7ex}{$\sim$}}\,}   
\newfont{\Sc}{eusm10}

\begin{document}

\title{MOIRCS Deep Survey III: Active Galactic Nuclei in Massive Galaxies at $z=2-4$}

\author{T. Yamada, \altaffilmark{1} 
   M. Kajisawa, \altaffilmark{1} 
   M. Akiyama, \altaffilmark{1}
   T. Ichikawa, \altaffilmark{1}\\
   M. Konishi, \altaffilmark{2}
   T. Nishimura, \altaffilmark{2}
   K. Omata, \altaffilmark{2}
   R. Suzuki, \altaffilmark{2} \\
   I. Tanaka, \altaffilmark{2}
   C. Tokoku, \altaffilmark{1}
   Y. K. Uchimoto, \altaffilmark{3}
   T. Yoshikawa, \altaffilmark{1,2}
}
\email{yamada@astr.tohoku.ac.jp}

\altaffiltext{1}
{Astronomical Institute, Tohoku University, Aramaki,
Aoba-ku, Sendai, Miyagi, 980-8578}

\altaffiltext{2}
{Subaru Telescope, National Astronomical Observatory of
Japan, 650 North A'ohoku Place, 
Hilo, HI 96720, U.S.A}

\altaffiltext{3}
{Institute of Astronomy, University of Tokyo, 2-21-1 Osawa, Mitaka,
Tokyo, 181-0015}

\begin{abstract}
  We investigate the X-ray properties of the $K$-band-selected galaxies at redshift $2 < z < 4$ by using our deep near-infrared images obtained in the MOIRCS Deep Survey project and the published {\it Chandra} X-ray source catalog. 61 X-ray sources with the 2-10 keV luminosity $L_X = 10^{42}-10^{44}$ erg s$^{-1}$ are identified with the $K$-selected galaxies and we found that they are exclusively (90\%) associated with the massive objects with stellar mass larger than $10^{10.5}$ M$_\odot$. Our results are consistent with the idea that the M$_{\rm BH}$/M$_{\rm str}$ ratio of the galaxies at $z=2-4$ is similar to the present-day value. On the other hand, the AGN detection rate among the very massive galaxies with the stellar mass larger than $10^{11}$ M$_\odot$ is high, $33$\% (26/78). They are active objects in the sense that the black-hole mass accretion rate is $\approx 1$-50\% of the Eddington limit if they indeed have similar M$_{\rm BH}$/M$_{\rm str}$ ratio with those observed in the local universe. The active duration in the AGN duty cycle of the high-redshift massive galaxies seems large.
\end{abstract}

\keywords{galaxies: active --- galaxies: formation --- galaxies: evolution}
\section{Introduction}

 It is now widely accepted that the most of the massive galaxies have their central super-massive black holes (SMBHs) (e.g., Kormendy \& Richstone 1995). The tight correlation between the mass of SMBH, M$_{\rm BH}$, and the spheroid mass, M$_{\rm sph}$, or the velocity dispersion of the host is observed for both normal and active galaxies in the nearby universe (Magorrian {\it et al.} 1998; Ferrarese \& Merrit 2000; McLure \& Dunlop 2001). The correlation suggests that the formation of galaxies and their SMBH are closely related and probably have the causal or even coeval connection. The correlation may result from (i) mass accretion to the BH due to the star-formation in galaxy spheroids through some physical process such as the stochastic kinematic viscosity by supernovae or by the radiation drag mechanism (Yamada 1994; Wada \& Norman 2002; Kawakatsu \& Umemura 2002), or (ii) termination or suppression of star-formation by the feedback of strong nuclear activity (Hopkins {\it et al.} 2006). The true origin of the relationship is, however, still far from unveiled especially in observations.

 In order to see how the formation of galaxies and their SMBH are related, and to understand the origins of the M$_{\rm BH}$-M$_{\rm sph}$ correlation, it is needed to study the relationship among stellar mass, star-formation, and active galactic nuclei (AGN) activities of the galaxies at high-redshift when they are majorly developing their stars, structures, and central black holes. It is also very interesting to see whether such relationship has always been hold or not during the hierarchical growth of massive galaxies (Haehnelt \& Rees 1993; Haehnelt \& Kauffmann 2000; Granato {\it et al.} 2004; Hopkins {\it et al.} 2006).  At $z$=0, AGN are observed exclusively in the massive galaxies with the stellar mass larger than $\sim 10^{10}$ M$_\odot$ (Kauffmann {\it et al.} 2003). Is this also true at high redshift, or do the progenitors of the present-day massive galaxies already have the SMBH as massive as they have now? Indeed, some authors claimed that the M$_{\rm BH}$/M$_{\rm sph}$ ratio increases with increasing redshift (McLure {\it et al.} 2006), in which case SMBH grow earlier than the formation of the entire stars in massive galaxies.

 It is, however, difficult to directly observe the M$_{\rm BH}$-M$_{\rm sph}$ correlation at high redshift. The techniques to evaluate the BH mass of normal galaxies cannot be used for the distant objects. For active galaxies at high redshift, massive BHs are likely to have been formed in luminous quasars (Turner 1991) and their virial mass can be measured by using the velocity width and the size of the broad-line regions estimated from the quasar luminosity (Kaspi {\it et al.} 2000; McLure \& Dunlop 2001). They are in the range of 10$^8$-10$^{10}$ M$_\odot$ for the luminous quasars at $z > 2$ (Shen {\it et al.} 2008).  On the other hand, however, it is difficult to directly observe the properties of the host galaxies, such as morphology, stellar mass, and velocity dispersions, due to the pollution of the light from the very bright nuclei. Adaptive optics imaging of distant quasars succeeded to detect host galaxies only for a few objects (e.g., Croom {\it et al.} 2004). Akiyama (2005) investigated the host galaxies of X-ray-selected AGN which are less luminous than quasars on the rest-frame UV images and detected extended components for the half of the samples (17/31), but their stellar mass is only weakly constrained due to the lack of the deep near-infrared data.  McLure {\it et al.} (2006) analyzed the sample of 3C RR radio-loud objects and argue that the M$_{\rm BH}$-M$_{\rm sph}$ ratio is larger at high redshift, $\sim 5\times$ of the local at $z \sim 2$. The line-width of the emission lines in radio-loud objects, however, may be affected by the gas kinematics due to the radio jet activity, which may cause possible uncertainty in BH mass estimation.

 Molecular gas, which is related to star-formation but not directly to the BH activity, is a reliable probe to see the host galaxies of luminous quasars (Yamada 1994, Miolino {\it et al.} 2007 and reference therein). Interestingly, more than a few luminous high-redshift quasars are detected in CO lines and the evidence of massive ($> 10^{11}$M$_\odot$) host galaxies of the quasars at $z$=2-6 were reported (e.g., Ohta {\it et al.} 1996; Miolino {\it et al.} 2007).  Evidence of the growing SMBH is also observed for high-redshift star-forming galaxies. Alexander {\it et al.} (2005; 2006) have revealed that $\sim 75$\% of the sub-mm-selected galaxies (SMGs) show AGN activities and the majority of the active nuclei seem to be heavily obscured by dust. Daddi {\it et al.} (2007) also studied the mid-infrared excess galaxies at $z \sim2$ and found the evidence of obscured Compton-thick AGN among them by the stacked X-ray data for the sample. These observations suggest that the formation of SMBH is indeed related with the formation of massive galaxies in their infrared  ultra luminous (ULIRG) phase. Time-lag between starburst and AGN activity may results in showing the high or low M$_{\rm BH}$-M$_{\rm sph}$ ratio in quasar and ULIRG phase (Alexander {\it et al.} 2008).

 In this paper, we study the general X-ray properties of the massive and the less massive galaxies at $z=2$-4 using the new extremely deep  near-infrared (NIR) imaging data obtained in the MOIRCS Deep Survey (MODS, see Section 2) to investigate the AGN activity among the stellar-mass based sample.  Previously, only the X-ray fraction among the NIR-selected sources at $z \sim 2$ have been reported (Reddy {\it et al.} 2005; Papovich {\it et al.} 2006) and the detailed properties of the X-ray sources associated with the galaxies at such higher redshift has not been studied well. At $z=2-4$, the X-ray sources detected even in the deepest observation seems dominated by AGN population with the X-ray luminosity $L_X$ (2-10 keV)= 10$^{42}$-10$^{44}$ erg s$^{-1}$, which corresponds to the luminosity of the luminous Seyfert galaxies in the local universe (Green {\it et al.} 1992; Akiyama 2005). While they may not be at the peak of the AGN/quasar activity in each individual galaxy, the lower-limit values of the existing BH mass can still be evaluated if we assume the Eddington limit in their luminosity. On the other hand, given the hypothetical form of the redshift evolution of the M$_{\rm BH}$/M$_{\rm sph}$ ratio, we may discuss the AGN activity along the stellar mass for the galaxies at such high redshift.

 In Section 2, we introduce the NIR and the X-ray data and the sample we use, and describe the method to obtain the various properties of the galaxies, such as stellar mass, reddening, star-formation rate. We show the X-ray properties of these samples in Section 3 and discuss the AGN activity among the high-redshift galaxies in Section 4. The cosmological parameters used here are canonical values after WMAP, namely, $\Omega_0$=0.3, $\Omega_\Lambda=0.7$, and $H_0$=70 kms$^{-1}$Mpc$^{-1}$. NIR magnitude values are in Vega system (Oke \& Gunn 1983; Fukugita et al 1996).

\section{Data and Samples}

 In order to study the relationship between AGN activities and galaxy properties at high redshift, we constructed a sample of $K$-band selected galaxies between $z$=2 and 4 from the MOIRCS Deep Survey (MODS, Kajisawa {\it et al.} 2006; Ichikawa {\it et al.} 2007). 

 By using Multi-Object InfraRed Camera and Spectrograph (MOIRCS, Ichikawa {\it et al.} 2006; Suzuki {\it et al.} 2008) equipped with Subaru 8.2m telescope, we obtained deep $J$, $H$, and $Ks$ band images at the sky area in the fields of Great Observatories Origins Deep Survey field North (GOODS-N).  We cover the area by the four fields of views (FoV) of MOIRCS, referred as GT1, GT2, GT3, and  GT4, from north-east to south west. Full description of MODS result will be published in the separated papers (Kajisawa {\it et al.} in preparation).  GT2, which includes the Hubble Deep Field North (HDF-N) is also the ultra-deep field of MODS survey where more than 28h science integrations were obtained in both $J$ and $Ks$  bands. The total sky area (GT1-GT4) used in the analysis of this paper is 103.3 arcmin$^2$. The basic properties of the data obtained and reduced by June, 2008 are summarized in Table 1. 

\begin{table}
\begin{center}
\caption{Summary of the NIR observation\label{tab1}}
\begin{tabular}{ccccccl}
\tableline\tableline
Field & Filter & FWHM (ch1) & Depth (ch1)\tablenotemark{a} & FWHM (ch2) & Depth (ch2)\tablenotemark{a} &  Exp. Time \\
      &        & arcsec   &  mag.        & arcsec      & mag.        & hours  \\
\tableline

GT1   &   J  &   0.59   &  24.9  &   0.59 &    24.8 &    8.0  \\
      &   H  &   0.58   &  23.8  &   0.59 &    23.7 &    2.5  \\
      &   K  &   0.58   &  23.7  &   0.53 &    23.8 &    8.3  \\
GT2   &   J  &   0.48   &  25.7  &   0.49 &    25.6 &    28.2 \\
      &   H  &   0.46   &  24.4  &   0.46 &    24.2 &    5.7  \\
      &   K  &   0.45   &  24.7  &   0.46 &    24.6 &    28.0 \\
GT3   &   J  &   0.57   &  24.9  &   0.58 &    24.8 &    6.3  \\
      &   H  &   0.55   &  23.7  &   0.55 &    23.7 &    3.2  \\
      &   K  &   0.59   &  23.8  &   0.60 &    23.7 &    10.7 \\
GT4   &   J  &   0.58   &  24.8  &   0.59 &    24.8 &    9.1  \\
      &   H  &   0.58   &  23.8  &   0.59 &    23.8 &    4.3  \\
      &   K  &   0.59   &  23.6  &   0.60 &    23.7 &    9.8  \\
         
\tableline
\end{tabular}
\tablenotetext{a}{The 3$\sigma$ limits given in Vega magnitude for the aperture with the diameter of 2xFWHM of the average point source profile. As MOIRCS has the two equivalent optical channels, we obtained the detection limits separately for each channel.}
\end{center}
\end{table}

 In MODS, more than 7000 $K$-band-selected galaxies are detected down to $K \sim 23.6$ (3$\sigma$) at GT1-GT4 and 24.6 in the deepest GT2 field. The MOIRCS images are carefully aligned to the optical images obtained by Advanced Camera for Survey (ACS)  on Hubble Space Telescope (HST), as well as the infrared images taken with IRAC and MIPS instruments on Spitzer Space Telescope. ACS and MOIRCS $J$ and $H$ images are smoothed to be matched with the $Ks$-band images and the colors of the objects are obtained by the apertures with the diameter of 1.2 arcsec.  For the IRAC data, we first measured the flux in apertures in 3.16, 2.78, 3.43, and 3.47-arcsec diameter in 3.6, 4.5, 5.8, and 8$\mu$m, and then made aperture correction by using the light profile of the $Ks$-band image smoothed to be matched with each IRAC band. The multi-band photometric data are then fitted by the galaxy evolutionary models of Bruzual \& Charlot (2003) so that we obtained the photometric redshift, stellar mass, reddening, and parameters of the star-formation histories such as ages, star-formation-rate, and the decaying parameter $\tau$ by the similar method described in Kajisawa \& Yamada (2005).  Kajisawa \& Yamada (2005) carefully investigated the uncertainty in the evaluation of the stellar mass and found that the stellar mass of the most of the $K$-selected galaxies can be determined well for the given SED while there may be degeneracy in other parameters such as age, extinction, and star-formation-decaying time scale. Kajisawa {\it et al.} (in preparation) also studied the uncertainty of the stellar mass estimation of the current sample to find that the $1\sigma$ uncertainty is  0.1 dex for M$_{\rm str}$=$10^{10.5}-10^{11}$M$_\odot$ and 0.2-0.3 dex for M$_{\rm str}$=$10^{9.5}-10^{10.5}$M$_\odot$.  In this paper, we used the models with Salpeter initial mass function (IMF).  Reduction by the factor of two in the stellar mass values is needed to compare with those obtained with the Chabrier-like IMF (Chabrier {\it et al.} 2003).

 The X-ray source catalog in the GOODS-N field is taken from Alexander {\it et al.} (2003) who conducted 2M-sec observations with the X-ray satellite {\it Chandra} at the field. The on-axis sensitivity limit is $\approx 2.5 \times 10^{-17}$ erg s$^{-1}$ at 0.5-2 keV (soft) band, $\approx 1.4 \times 10^{-16}$ erg s$^{-1}$ at the 2-8 keV (hard) band, and $\approx 1.0 \times 10^{-16}$ erg s$^{-1}$ at the 0.5-8 keV (total) band, respectively. Their exposure maps show that the data is fairly homogeneous over the MODS GT1-GT4 fields. The median positional determination accuracy is $\approx 0.3$ arcsec (Alexander {\it et al.} 2003).   

 We matched the Chandra X-ray source positions in their catalog (Table 3 in Alexander {\it et al.} 2003) with the $Ks$-band sources by the following procedure. We first identify the pixel on the $Ks$-band image at the X-ray source position in the catalog and then find the $Ks$-band selected object to which the pixel belongs. If there is no $Ks$-band object at the pixel, we then extend the search pixels up to three (0.35 arcsec) to find the $Ks$-band counterpart. In fact, the most of the Chandra sources can be identified with the $Ks$-band sources in the first step thanks to the good positional accuracy of the Chandra data and the depth of the infrared data. While the depth of NIR images at the field GT2 is significantly deeper than other fields, most of the X-ray sources but a few are identified with the $Ks$-band sources above $K \sim 23$.

 For the current purpose, $K$-selected sample within the redshift range, $2 < z < 4$ are constructed. $Ks$-band light samples $\sim 4000-7000$\AA\ at rest frame at this redshift range. We primary use the spectroscopic redshift if available (Wirth {\it et al.} 2004; Reddy {\it et al.} 2006; Barger {\it et al.} 2008) and photometric ones if not.  In total, over GT1-GT4, 2621 galaxies are selected as the sample of which 217 objects have spectroscopic redshift. Among them, 1374 galaxies are brighter than $K$=23, which is the conservative completeness limit over GT1-GT4. 79 objects have the stellar mass larger than 10$^{11}$ M$_\odot$ and 219 objects have 10$^{10.5}$ M$_\odot$$<$ M$_{\rm str}$ $<$ 10$^{11}$ M$_\odot$. The accuracy of the photometric redshift, evaluated by using the spectroscopic sample, is $\delta$$z$/$(1+z)$=-0.009$\pm$0.093 with 9$\%$ outlier that has $\|$$\delta$$z$/$(1+z)$$\|$$>0.5$ and $\delta$$z$/$(1+z)$=-0.071$\pm$0.115 with 5$\%$ (1 source) outlier for the $K$-selected galaxies and those associated with the X-ray sources, respectively. The redshift distribution at $2 < z < 4$ is shown in Fig.1.

\section{Results}

 Fig.2 shows the stellar mass versus $J-K$ color of the 2621 objects. Galaxies with $K <23$, above the completeness limit over GT1-GT4, are shown by the large dots and the fainter ones are by the small dots. As previously pointed out (e.g., Kajisawa \& Yamada 2005; 2006), the most massive galaxies tend to have the redder colors. At M$_{\rm str}$ $> 3 \times 10^{10}$ M$_\odot$, more than 80\% of the galaxies have $J-K > 2$.

 The galaxies identified with the Chandra sources are also shown in Fig.2 and the X-ray identified fractions of the samples are listed in Table 2.  In total, 61 total-band (0.5-8 keV) X-ray sources, or  52 hard-band (2-8 keV) X-ray sources are identified with the $K$-band sources. We found that 21 of 61, or 19 of 52 objects have the spectroscopic redshift values between $z$=2 and 4. Since the majority of the X-ray sources are detected in the hard energy band, 2-8 keV, which corresponds to 6-24 keV and 10-40 keV at $z$=2 and 4, respectively, these X-ray sources are likely to be AGNs. Indeed, their luminosity range, $10^{42-44}$ erg s$^{-1}$, is consistent with those of the luminous Seyfert galaxies or faint quasars in nearby universe. While active star-formation also produces luminous X-ray emission, the expected luminosity is more than an order of magnitude lower than the measured values. The empirical fit by Ranalli {\it et al.} (2003) predicts $L_X$(2-10 keV) = $5.0 \times 10^{39}$ erg s$^{-1}$ for the star-formation rate (SFR) of 1 M$_\odot$ yr$^{-1}$. As shown below, the largest SFR estimated for the current sample is 1000 M$\odot$ yr$^{-1}$ that may produce the X-ray luminosity of $5.0 \times 10^{42}$ erg s$^{-1}$, which is about the lower bound of our sample (see Fig.8 below). We conclude that the observed X-ray emission is associated with the AGN activity without serious contamination by star-formation.

 We first note that the most of the identified X-ray sources are associated with massive galaxies. Among the 61 total-band sources, 90\% (55/61) are associated with the galaxies with M$_{\rm str}$ $>10^{10.5}$ M$_\odot$ and 48\% (29/61) are those with M$_{\rm str}$ $>$ $10^{11}$ M$_\odot$. Table 2 also summarizes these numbers.  On the other hand, for the massive galaxies with M$_{\rm str} > 10^{11}$ M$_\odot$, the fraction of the X-ray detection is notably high, 33\% (26/78) and 29\% (23/78) for the total-band and the hard-band sources, respectively. These values can be interpreted as the lower limit as we ignored the attenuation  to the X-ray emission; we may missed very Compton-thick objects with absorption larger than N$_{\rm H}$ $\approx$ $10^{24}$ cm$^{-2}$.

 We checked by the several methods that the NIR light of the objects is indeed stellar origins and not heavily contaminated by the AGN component and the evaluation of the stellar mass based on NIR luminosity is not significantly affected by the AGN light. 
 If we adopt the typical X-ray to optical flux-density (per unit frequency) ratio for broad-line AGN, $f_{\nu}$(20 keV)/$f_{\nu}$(0.5$\mu$m) $\sim 10^{-4}$-$10^{-5}$ (Richards {\it et al.} 2006), the contribution of the AGN light in $Ks$-band, namely rest-frame optical band, is estimated to be 19.5-22 mag for the sources with $f_{\rm X}$(2-8 keV)=$10^{-15}$ erg s$^{-1}$ cm$^{-2}$ without extinction at the optical wavelength. While the most of the X-ray sources in our sample are fainter than this flux range and their host-galaxy $K$-band magnitudes are in the range of $K$=19-22.5 mag, the contribution of the AGN light in $K$ band may not be negligible for more than a few cases. 

 We therefore carefully examined the colors and the optical-mid-infrared spectral energy distributions (SED) of the $K$-band sources which have the X-ray counterpart.  Donley {\it et al.} (2007) examined the properties of the Spitzer IRAC sources in GOODS-N which have 'power-law' SEDs and found about a half of them are detected in X-ray to show the AGN activity. We checked how many $K$-selected sources  may be classified with the 'power-law' sources and found that 13 $K$-selected sources are matched with such SED in Donley {\it et al.} (2007) and of them, 9 are detected by {\it Chandra}. Other four non-Xray objects may host hidden very Compton-thick AGN, which make the AGN fraction of the sample even larger. Thus about 15\% (9/61) of the X-ray-identified $K$-selected galaxies the NIR light may be contaminated by the hot-dust component. In our own photometry, 7 of the 9 'power-law' sources (detected in X-ray) indeed have the SEDs monotonically increased in AB magnitude toward the longer wavelength. 
 However, the rest (85\%, 52/61) of the sample can be well fitted by the stellar-population SEDs showing the turn off near the rest-frame '1.6$\mu$m bump', which supports that our stellar-mass evaluation for these objects is robust. Furthermore, even for the 9 'power-law' sources, we found that our stellar-mass estimation may be reasonable. We made the similar SED fitting process with regarding the $K$ and the IRAC magnitudes as the upper limits, allowing the contamination by the AGN light, but the evaluated stellar mass values do not change very much with a few exceptions. Fig.3 compare the stellar mass obtained by the SED fitting to the all measured photometric values and those with the upper-limit substitution in $K$ and the IRAC bands. 
 We also show the three examples of the SED fitting results in Fig.4a-4c. Fig.4a shows the very typical case which has the middle-range $\chi^2$ value in the fitting among the sample. The magenta line is the best-fitted models using the all photometric data points and the blue line is that using the $K$ and IRAC-band data (open circles) as the upper limit values. In this case, clear turn off around the 5.8$\mu$m-band (IRAC ch3) data, which corresponds to the 1.6$\mu$m-bump feature in the stellar light, is appeared. Fig.4b shows the case of a relatively red objects, which has the $\chi^2$ value at $\sim16\%$ (10th among the 61 objects) from the worst. While the 8$\mu$m data is deviated from the stellar model, which shows the evidence of hot dust component, the data at the wavelength shorter than 5.8$\mu$m are well fitted by the stellar model including the Balmer jump feature near the $H$ band. In this case, the difference in stellar mass from the two different fitting procedures is $\sim 50\%$ , which is one of the largest deviation among the sample (see Fig.3). Fig.4c shows the case of a 'power-law' object, which indeed shows the monotonic rising of the flux toward the longer wavelength. Even in this case, we see clear jump of the SED below the $J$ band, which is very likely to the Balmer jump in the stellar spectrum. The obtained stellar mass is almost same for the two different fitting procedures while the other parameters of the models (age, extinction) are different.

 We also found that the $Ks$-band light distributions show the extended profile and not dominated by the point source of the AGN components. Fig. 5 shows the FWHM of the $Ks$-band images of the 61 X-ray sources. Majority ($\gtsim 70\%$) show extended profiles. At the bright end, or in the sharpest image, GT2, the fraction is higher. The 'power-law' sources are marked with the squares and we see that their $Ks$-band light distributions are also extended in many cases.
 From these above, we conclude that the observed NIR light are not seriously contaminated by the AGN light and the obtained stellar mass values are as robust as for the non-X-ray galaxies.

\begin{table}
\begin{center}
\caption{Summary of the $K$-band and X-ray Sources\label{tab1}}
\begin{tabular}{cccc}
\tableline\tableline
      & All &  M$_{\rm str}$ $>$ $10^{11}$ M$_\odot$ &  $10^{10.5}$ M$_\odot$ $<$ M$_{\rm str}$ $<$ $10^{11}$ M$_\odot$  \\

\tableline

$K$-band Sources with $2 < z < 4$  &  2621 &  79  &  219 \\
Hard-band X-ray Sources            &   52  &  23  &  24  \\
Total-band X-ray Sources           &   61  &  26  &  29  \\

\tableline
\end{tabular}
\end{center}
\end{table}

%
 Fig.6 shows the X-ray luminosities of the sources as the function of the stellar mass of the host galaxies.  We obtained the X-ray luminosity at 2-10 keV from the hard-band flux adopting the relation $f_\nu \sim \nu^{-0.7}$ to obtain the K-correction factor. While the soft band (0.5-2 keV) is closer to the rest-frame 2-10 keV energy range for the objects at $z=2-4$, the hard-band flux is less affected by the extinction and we may obtain the X-ray luminosities which are closer to the intrinsic values. For those which are not detected significantly in the hard-band, we used the total-band energy flux. The galaxies between $z=2$ and 3 are shown by the blue dots and the galaxies at $z=3$-4 by the red dots. There is no clear trend in Fig.6 and thus the stellar mass and the X-ray luminosity does not seem to strongly correlate with each other.

 In Fig.7, we plotted the 'specific AGN activity', namely the X-ray luminosity divided by the host stellar mass, against their host stellar mass. Note that this is just a different presentation of the data shown in Fig.6. As expected from Fig.6, notable trend that more stellar massive galaxies have smaller specific AGN activity is observed. If the M$_{\rm BH}$/M$_{\rm str}$ ratio is the same among the galaxies and the BHs are similarly active (in mass accretion), the specific AGN luminosity is to be constant. For the guide, we plotted the case of the local  M$_{\rm BH}$/M$_{\rm sph}$ ratio and the 10$\%$ Eddington accretion rate by the dotted line and 100$\%$ and 1$\%$ cases by the upper and lower dashed lines, respectively. Fig.7 shows that the specific AGN activity is smaller than the 10$\%$-Eddington value for the massive galaxies with M$_{\rm str}$ $> 10^{11}$ M$_\odot$ and larger for the less massive galaxies near M$_{\rm str}$ $= 10^{10.5}$ M$_\odot$ . This indicates that BHs in massive galaxies are relatively inactive, or, that more massive BHs are formed in relatively less massive galaxies and they have the similar activities. We discuss these results further in Sec. 4. 

 It is also possible that attenuation of the X-ray emission in massive galaxies causes such a trend. In order to see whether this is the case, we checked the {\it Spitzer} MIPS 24$\mu$m flux data of the sample since the objects with luminous infrared emission may host such obscured AGN. The objects with large circles in Fig.7 are those detected in MIPS 24$\mu$m band above $\sim 80$$\mu$Jy. The MIPS detection fraction among the massive objects, M$_{\rm str}$ $> 5 \times 10^{10}$ M$_\odot$, is 39$\%$ (7/18) for those with the relatively large specific AGN activity (objects above the dotted line in Fig.7) and 52\% (13/25) for the rest. Thus the 24$\mu$m detection rate is similar for the two samples, which does not support that the low specific AGN activity is due to the X-ray attenuation in dust-rich environment. 

 Fig.8a and 8b show the distribution of the X-ray luminosity and the star-formation rate (SFR) of the galaxies obtained by the SED fitting. We use the 'instantaneous' SFR in the left panel (Fig.8a) and the SFR averaged over the past 300 Myrs in the right (Fig.8b). The SFR ranges from a few to 1000 M$\odot$ yr$^{-1}$ and the specific SFR, namely the SFR divided by their stellar mass, ranges from 10$^{-11}$ to 10$^{-7}$ yr$^{-1}$. We do not see any clear correlation between $L_{\rm X}$ (2-10 keV) and SFR over 1-1000 M$_\odot$yr$^{-1}$. If the star-formation in each galaxy and the mass accretion to its BHs are closely correlated and exactly coeval all the time, there should be more tight correlation between $L_{\rm X}$ and SFR, while a large uncertain here is how much fraction of the SFR is indeed directly related to the formation of the spheroid component. As also inferred from Fig.8, the specific AGN activity and the specific SFR do not show tight correlation.  The dotted lines in the figures show the expected X-ray luminosity due to the star-formation activity predicted from the empirical fit to the local galaxies by Ranalli {\it et al.} (2003). The recent models studied by Mas-Hesse {\it et al.} (2008) give the consistent results. We see the observed X-ray luminosities are an order-of-magnitude more luminous than the prediction, which implies that they are due to the AGN activities and not seriously contaminated even by the very intense star formation.

 In Fig.9, we also show the distribution of the $K$-selected as well as the X-ray-identified objects along the color excess values, E(B-V) and the average age. We adopted the Calzetti law for the dust extinction in the SED fitting. The X-ray-identified sources are distributed over the parameter plane and no very strong trend is recognized. Ages and E(B-V) values, however, may be degenerate in the SED fitting, which possibly dilutes any intrinsic trends. There is a hint that the X-ray fraction is slightly high at the middle range of E(B-V) between 0.3 and 0.6 for the massive galaxies but at the same time we note that the most reddened objects have few X-ray counterparts.

\section{Discussion}

\subsection{X-ray Sources in Distant Red Galaxies}

 Our results in the last section largely owe to the photometric redshift estimation. Although the 40\% of the X-ray sources have the secure spectroscopic redshift values and the comparison between photometric and spectroscopic redshift for the $K$-selected galaxies is also  reasonable (Kajisawa and Yamada 2005), it is worth analyzing the similar sample based on the simple color cut avoiding the uncertainty of the detailed photometric redshift measurements.

 For the purpose, we defined the sample of {\it Distant Red Galaxies} (DRGs), which have $J-K$  $> 2.3$ (Franx {\it et al.} 2003; van Dokkum {\it et al.} 2003).  DRGs are supposed to be massive galaxies ($M_* \sim 10^{11} M_{\odot}$) at $2 \ltsim z \ltsim 4$ (Franx {\it et al.} 2003; F\"{o}rster Schreiber {\it et al.} 2004; Reddy {\it et al.} 2005; Kajisawa {\it et al.} 2006) or dusty galaxies at intermediate redshift, $1 \ltsim z \ltsim 2$ (Webb {\it et al.} 2006; Conselice {\it et al.} 2007; Lane {\it et al.} 2007).  While more than a half of the relatively bright DRGs with $K < 20$ have $z < 2$, the fainter DRGs are likely to be dominated by the galaxies at high redshift (Kajisawa {\it et al.} 2006). This simple color cut is thus useful to make a sample of massive galaxies at $z \gtsim 2$. Above $K=22.8$, the completeness limit of the color selection over GT1-GT4 fields, 182 DRGs are identified.

 Fig.10 shows the number counts of the whole DRGs as well as DRGs with the X-ray counterparts. Their color-magnitude distribution is plotted in Fig.11. The vertical dotted line in Fig.10 shows the detection limit for DRG, $K$=22.8. Note that the decrease of the number density of DRGs is seen at around $K$=22, which is more rapid than that of the entire $K$-band selected objects. This indicates that there are less red low mass galaxies at high redshift (Kajisawa {\it et al.} 2006). The X-ray DRGs are peaked at $K \sim 20.5$ mag, which corresponds to the massive galaxies with M$_{\rm str}$ $\sim$ $10^{11}$ M$_\odot$ at $z \sim 3$. At $K < 21$, 30-35\% of the DRGs (19/65 for the hard band, 23/65 for the total band) are identified with the X-ray sources while it decreases to 21\% at $K < 22.8$. 

 These analysis for the sample of DRGs with the simple color cuts thus support the trend discussed in the previous section, namely, a large X-ray fraction among the bright ($K<21$) or massive objects, and decrease of the X-ray fraction at the fainter magnitude.

\subsection{AGN Activity in Galaxies at $z=2$-4}

 Among the $Ks$-band selected galaxies at $z=2$-4, the X-ray emissions are exclusively associated with the massive objects, with M$_{\rm str}$ $> 10^{10.5}$ M$_\odot$. As the {\it Chandra} 2M-sec data is deep enough to constrain the presence of moderately-luminous AGN even at such high redshift, we may discuss the possible BH mass limits as a function of their host galaxy mass.

 The limiting X-ray flux $f$(2-8 keV)=$1.4 \times 10^{-16}$ erg s$^{-1}$ cm$^{-2}$ corresponds to $L_X$(2-10 keV) = $3 \times 10^{42}$ erg s$^{-1}$ to $1.4 \times 10^{43}$ erg s$^{-1}$ at $z=2$ to 4, which corresponds to the luminosity of the AGN such as luminous Seyfert galaxies in nearby universe.
 If we assume the bolometric fraction of the X-ray emission at 2-10 keV, $\epsilon_X$ and the ratio of the bolometric to the Eddington luminosity of the AGN, $R_{\rm edd}$, the observed X-ray luminosity can be connected to the black hole mass by the relationship, $$ L_X = 4 \pi G m_p c M_{\rm BH} \epsilon_X R_{\rm edd} $$ 
$$ = 2 \times 10^{43} (M_{\rm BH}/10^8 {\rm M}_\odot) (\epsilon_X/0.01) (R_{\rm edd}/0.1) erg s^{-1} $$.

 The absence of luminous AGN in the less massive galaxies indicates that the associated BH mass or the activity of the AGN is lower than the limit. Unfortunately from the current data we cannot discriminate which is the case; we need further spectroscopic observations to evaluate virial BH mass to break this degeneracy. Instead, we here discuss the two cases with rather extreme assumptions to see
  which the more natural interpretation is.

 We first consider the case that SMBH mass does not depend on the host stellar mass, namely similar massive (or less massive) BHs exist in massive or less massive host galaxies. In this case, the X-ray non-detection in less massive galaxies below M$_{\rm str}$ $\sim 3 \times 10^{10}$ M$_\odot$ is due to the lower activity (less mass accretion rate) of their BH. However, this is the opposite trend seen for the {\it X-ray detected} sample above $\sim 3 \times 10^{10}$ M$_\odot$ as shown in Fig.7 and thus not favorable; it will be strange that the activity of AGN falls off suddenly at $\sim 10^{10}$ M$_\odot$. 

  We then consider the second case that the M$_{\rm BH}$/M$_{\rm sph}$ ratio is constant and does not evolve from the local value, M$_{\rm BH}$/M$_{\rm sph}$$\approx 0.002$ and not depends on the host stellar mass. We also assume M$_{\rm str}$ $\approx$ M$_{\rm sph}$. In this case, the decrease of specific AGN activity among the X-ray {\it detected} sample above $\sim 10^{10}$ M$_\odot$ shown in Fig.7 can be interpreted as the decrease of relative mass accretion rate; we see that the Eddington ratio of SMBH in the host with M$_{str}$ $= 3 \times 10^{10}$ M$_\odot$ (or the corresponding M$_{BH}$ $= 6 \times 10^7$ M$_\odot$) is $R_{\rm edd} \approx 1$ and it becomes lower, $R_{\rm edd}$=0.1-0.01, for the more massive objects.  On the other hand, few X-ray AGN is detected for galaxies with the stellar mass less than 10$^{10}$ M$_\odot$ (Fig.2), which should host SMBH with M$_{\rm BH}$ $\sim 2 \times 10^7$ M$_\odot$ with the assumption above. This BH mass expected from the no-evolution hypothesis, which may be the upper limit values if we consider the lower spheroidal mass fraction, is comparable with the 'BH mass detection limit' for the current {\it Chandra} X-ray data, $\approx 2 \times 10^7$ M$_\odot$ and $\approx 1 \times 10^8$ M$_\odot$ for $R_{\rm edd}=0.1$ at $z=2$ and $z=4$, respectively, and $2 \times 10^6$ and $1 \times 10^7$ M$_\odot$ for $R_{\rm edd} = 1$.  
 Thus the latter interpretation seems more natural. Our results are at least consistent with that, at $z=2-4$, the galaxies with stellar mass with $\sim 10^{10}$ M$_\odot$ or less do not host the black hole which is significantly (i.e., more than an order of magnitude) heavier than that expected from the local relationship, i.e., M$_{\rm BH}$$\sim 10^7$ M$_\odot$ unless the accretion rate is very low, $R_{\rm edd} < 10^{-2}$.

 McLure {\it et al.} (2006) argue for the sample of 3C RR radio-loud objects that  M$_{\rm BH}$/M$_{\rm sph}$ ratio is an increasing function with redshift and the ratio is $\sim 5 \times$ larger at $z \sim 2$. These 3C RR galaxies are very massive, M$_{\rm str}$ $\sim 10^{11.5-12}$ M$_\odot$ and their BH mass is estimated to be M$_{\rm BH}$ $\sim 10^{9-10}$ M$_\odot$. If the trend is also true for the galaxies in our sample, the galaxies with M$_{\rm str}$ $\sim 10^{10}$ M$_\odot$ may have the BH with M$_{\rm BH}$ $\sim 10^{8}$ M$_\odot$, which is not favored in our results.

 The trend that the AGN activity seen only among the massive galaxies is very similar with that seen for the local galaxies. Kauffmann {\it et al.} (2003) investigated the host properties of the $\sim 20000$ AGNs among the $\sim 120000$ galaxies at $0.002 < z < 0.3$ and found that AGN with all luminosity reside almost exclusively in massive galaxies, with M$_{\rm str}$$\gtsim 10^{10}$M$_\odot$. While we have to consider the X-ray detection limits carefully, the current results are consistent with that AGN are observed only among the massive galaxies even at $z=2-4$. 

 On the other hand, a steep number-density evolution of massive galaxies from $z \sim 3$ to $z \sim 1$ is reported (Caputi {\it et al.} 2005). It is very likely only 10-30\% of the present-day massive ($>10^{11}$ M$\odot$) galaxies have been formed at $z \sim 3$ and if their progenitors already exist at such high redshift, they may be less massive galaxies which will be merged with other objects or forming more stars to grow. It is an interesting question whether such progenitors have already hosted the massive BHs which are observed in their present-day descendants or not. In our results, galaxies with smaller mass ($\ltsim$ a few $\times 10^{10}$ M$_\odot$) do not appear to host the BH as massive as those located in the present-day massive ($>10^{11}$ M$\odot$) galaxies. In this sense, it is unlikely that massive BH had already been formed in the less-massive progenitors of their present-day host galaxies. We may argue that massive BHs have been developed as the host galaxies have grown their stellar mass, and not significantly prior to that.

\subsection{Coeval Formation of BH and Galaxies}

 It is also surprising that a very large fraction ($\sim 33$\%) of the massive galaxies host the {\it active} AGN. If the M$_{\rm BH}$/M$_{\rm str}$ ratio for the massive galaxies does not differ from the local value, $\approx 0.002$, many of these AGN are active with the Eddington ratio $R_{\rm edd} > 0.01$ as we can see in Fig.7. While the observed X-ray luminosity and SED-fitting-based SFR are not strongly correlated and the BH formation (mass accretion) and the star formation seems not coeval exactly (Fig.8), it can be said that the active duration in the AGN duty cycle of the high-redshift massive galaxies is large. 

 At the same time, we already noted that the negative dependence of the specific AGN activity on the stellar mass is observed. We also investigated how the specific AGN activity depends on the averaged stellar age obtained in the SED fitting. The results are shown in Fig.12. While they are scattered out and no tight correlation exists, we see a broad trend that the AGN activity is high in the young host galaxies with the stellar age less than $10^8$ yr and relatively low in the galaxies which are older than $10^9$ yr; 8 of 11 galaxies with $T_{\rm AGE}$ $< 10^8$ yr have $R_{\rm edd} = 0.1-1$ and 9 of 14 galaxies  with $T_{\rm AGE}$ $> 10^9$ yr have $R_{\rm edd} = 0.01-0.1$.

 From these above, we may infer a possible scenario that the formation of the BHs in massive galaxies continues over the large fraction (i.e., $> 30\%$) of the period of their star formation although not instantaneously coeval. In the $K$-band-selected very massive ($\sim 10^{11}$ M$_\odot$) galaxies at $z=2-4$, the BH growth is still ongoing but as the star-formation activity peaked earlier in more massive galaxies,  as inferred from their relatively older age and lower specific SFR, their AGN activities may also be peaked earlier and only relatively lower-level activity is seen at the observed epoch.

\section{Summary}

 We investigated the X-ray properties of the $K$-band selected galaxies at $2 < z < 4$ with the stellar mass of 10$^9$-10$^{12}$ M$_\odot$ in 103 arcmin$^2$ in GOODS-N. $61$ X-ray sources, most of which must be AGN, are identified with the galaxies and we found that they are exclusively ($>$90$\%$) associated with the massive objects with the stellar mass $> 10^{10.5}$ M$_\odot$. On the other hand, a very large fraction, 30-35\% of the galaxies with stellar mass larger than 10$^{11}$ M$_\odot$ shows the AGN activity with the X-ray luminosity larger than $\sim 10^{42}$ erg s$^{-1}$. We also found that the specific AGN activity, namely the X-ray luminosity divided by the stellar mass is smaller in the more massive galaxies. The absence of luminous AGN in the less massive galaxies implies that the associated BH mass is smaller or the activity of the AGN is lower. Our results are consistent with the idea that the M$_{\rm BH}$/M$_{\rm str}$ ratio in $z=2-4$ massive and less massive galaxies is similar to the present-day value. Since the X-ray luminosity and the SFR are not correlated, the growth of BH and the star formation in the host seems not exactly coeval. At the same time, the large X-ray fraction of high-redshift massive galaxies also indicates that a significant fraction of the AGN activity continues over the comparable time scale of the star-formation in the host. 

\vspace*{5mm}

 We thank the staff of the Subaru Telescope for their assistance with the development and the observation of MOIRCS. This research is supported in part by the Grant-in-Aid 20450224 for Scientific Research of the Ministry of Education, Science, Culture, and Sports in Japan. Data analysis was in part carried out on common use data analysis computer system at the Astronomy Data Center, ADC, of the National Astronomical Observatory of Japan.

\clearpage



\figurenum{1}

\begin{figure}
\plotone{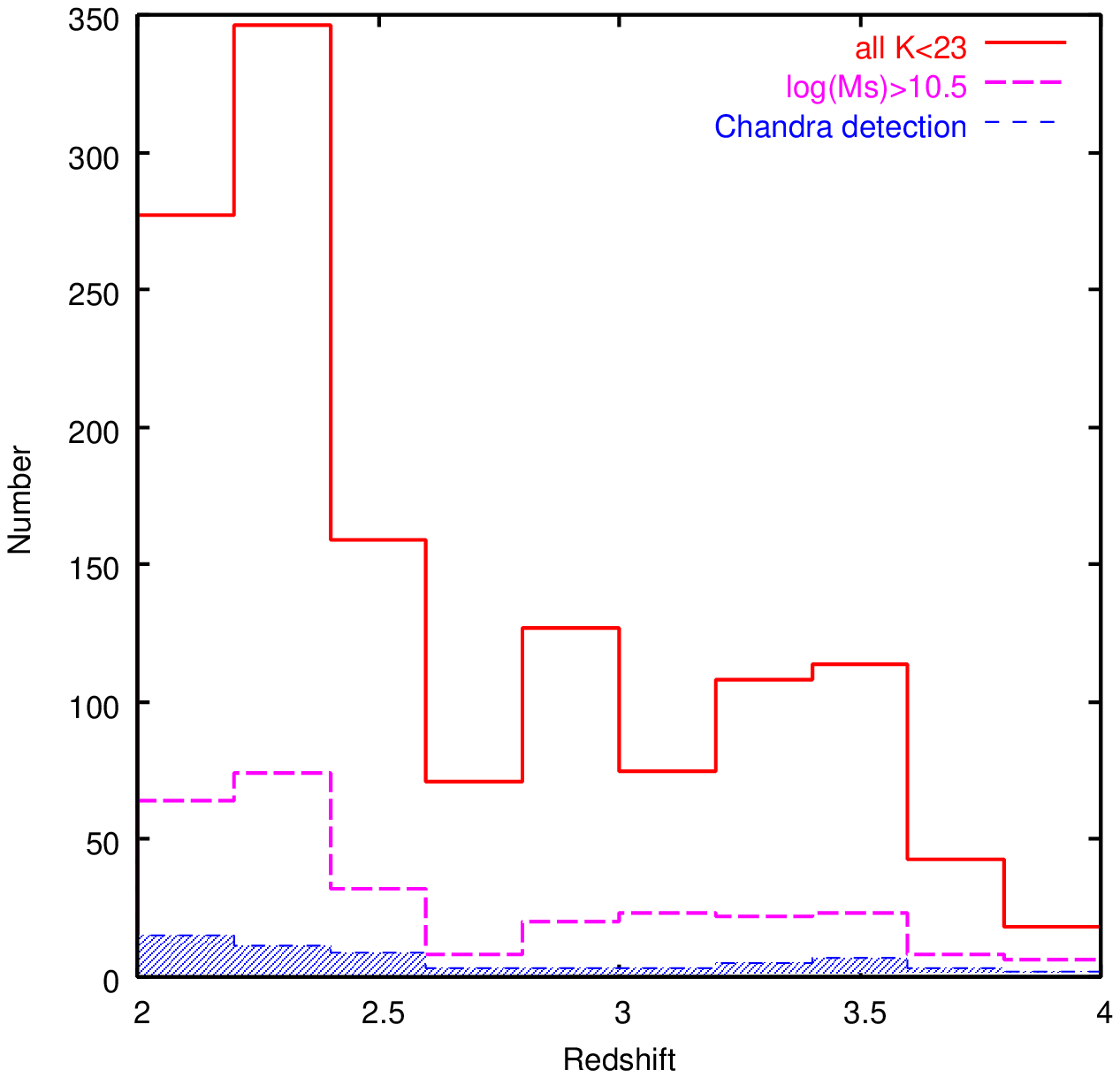}
\caption{The redshift distribution of the sample galaxies with $K$ $<23$. Those with stellar mass larger than 10$^{10.5}$ M$_\odot$ as well as those with X-ray detection are shown by the lower histograms.}
\label{fig1}
\end{figure}
%
\figurenum{2}
%
\begin{figure}
\epsscale{.80}
\plotone{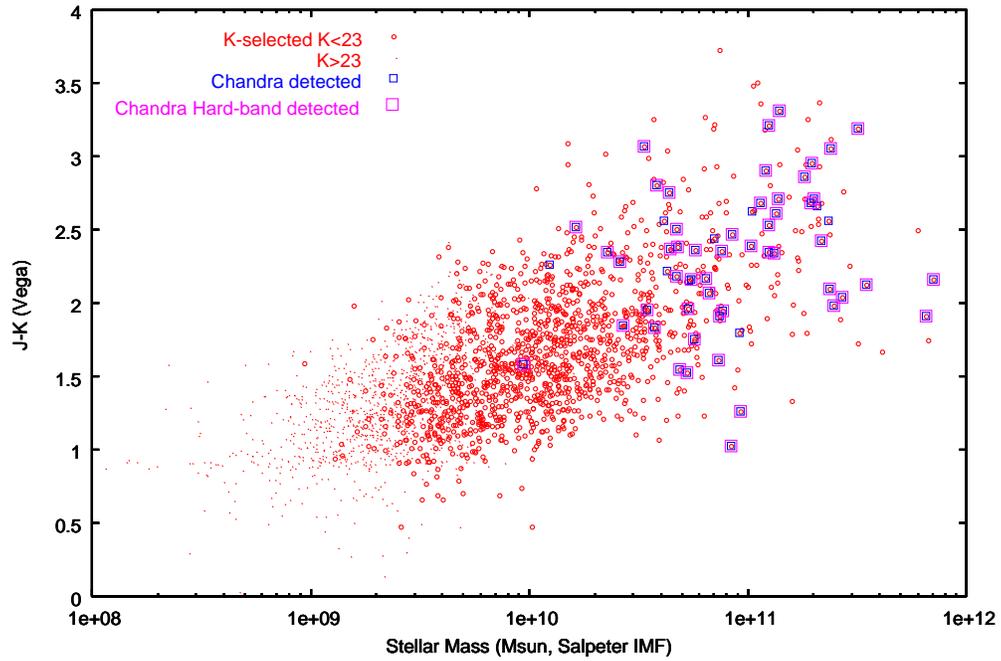}
\caption{The stellar mass and $J-K$ color of the galaxies in GT1-GT4 with photometric redshift $2 < z < 4$. Spectroscopic redshift are used if available. Galaxies with $K<23$ are shown by the large dots and the fainter ones by the small dots. Galaxies detected in the Chandra hard band and the total band are marked by the large and small squares, respectively.
}
\label{fig2}
\end{figure}
%
\figurenum{3}

\begin{figure}
\plotone{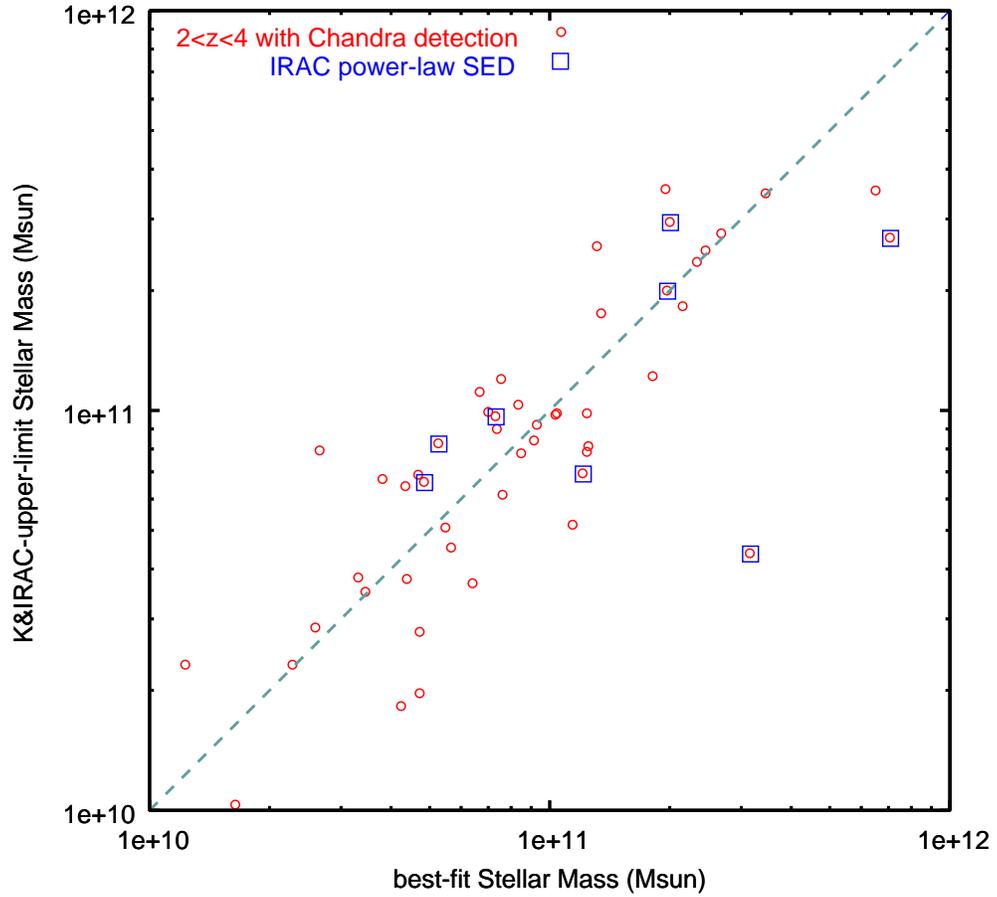}
\caption{Stellar mass obtained as the best-fitted value with the measured photometric data is compared with those obtained by treating the $K$ and the IRAC photometry as the upper limit values.}
\label{fig3}
\end{figure}
%
%
\figurenum{4}

\begin{figure}
\plotone{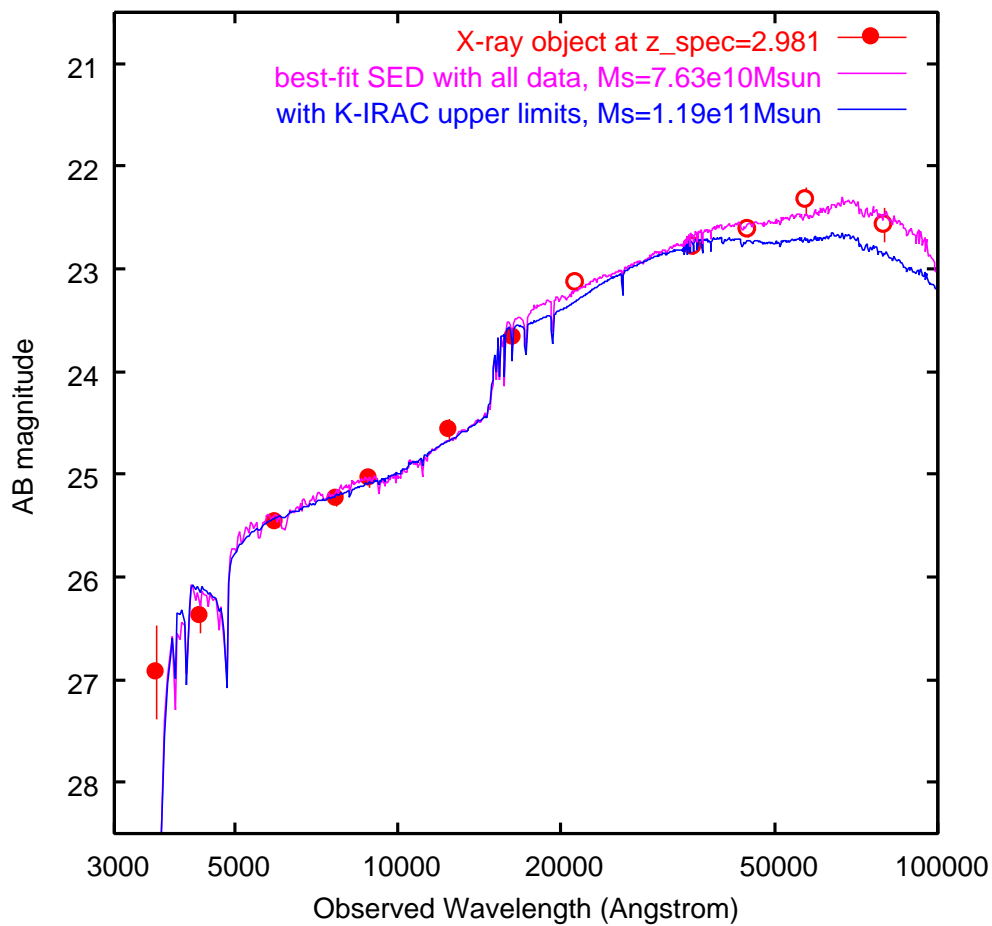}
\caption{Examples of the SED fitting results of the galaxies associated with the X-ray sources. Results of fitting with all the bands (magenta) and that using the K and IRAC bands as upper limits (blue) are shown. (a) A typical example with the $\chi$$^2$ value middle of the sample. (b) Example of objects with relatively red SEDs. The $\chi$$^2$ value is the 10th largest (16\%\ from the worst). (c) An example of the 'power-law' SED case. }
\label{fig4a}
\end{figure}
%
%
\figurenum{4}

\begin{figure}
\plotone{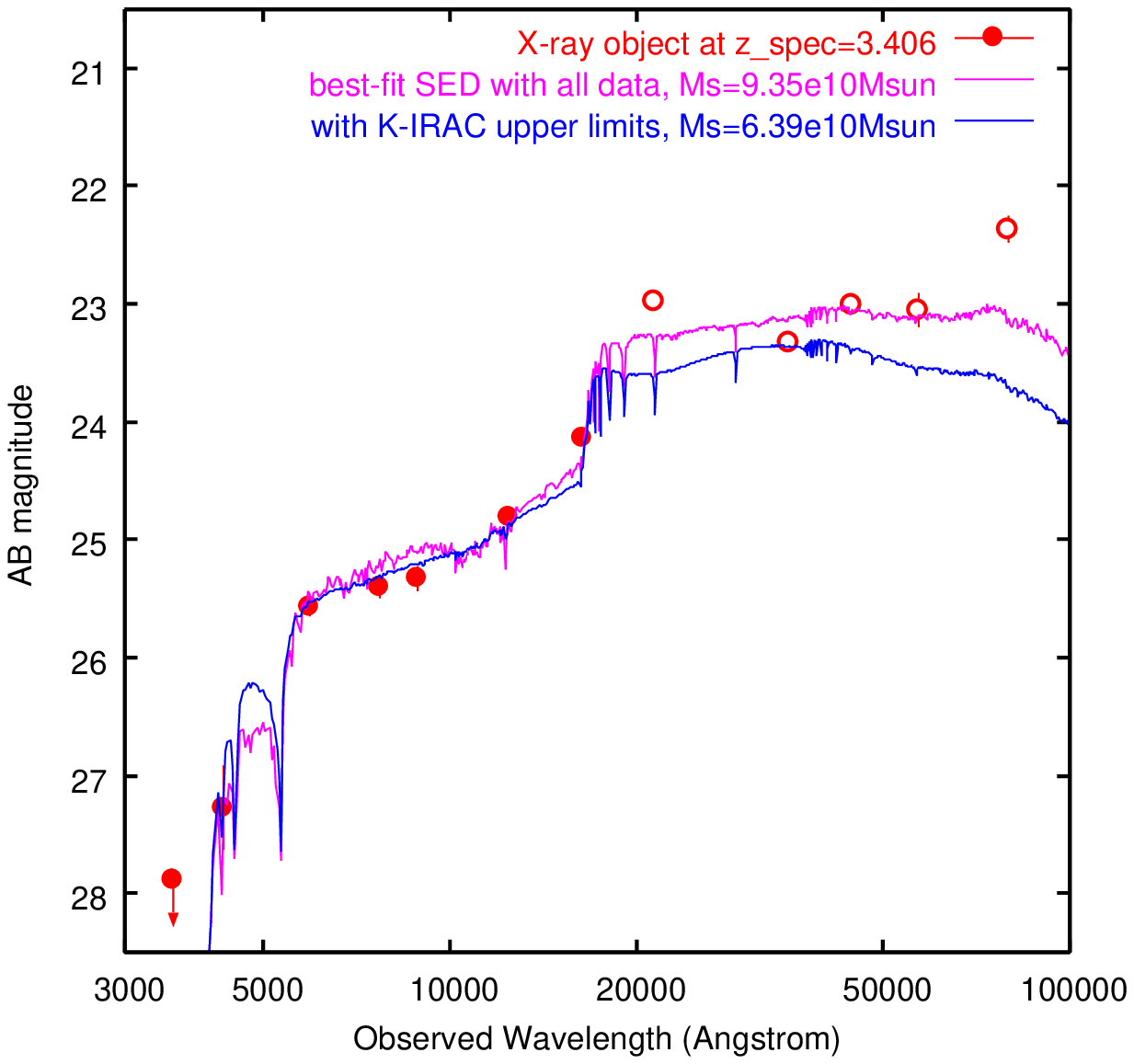}
\caption{ }
\label{fig4b}
\end{figure}

\figurenum{4}

\begin{figure}
\plotone{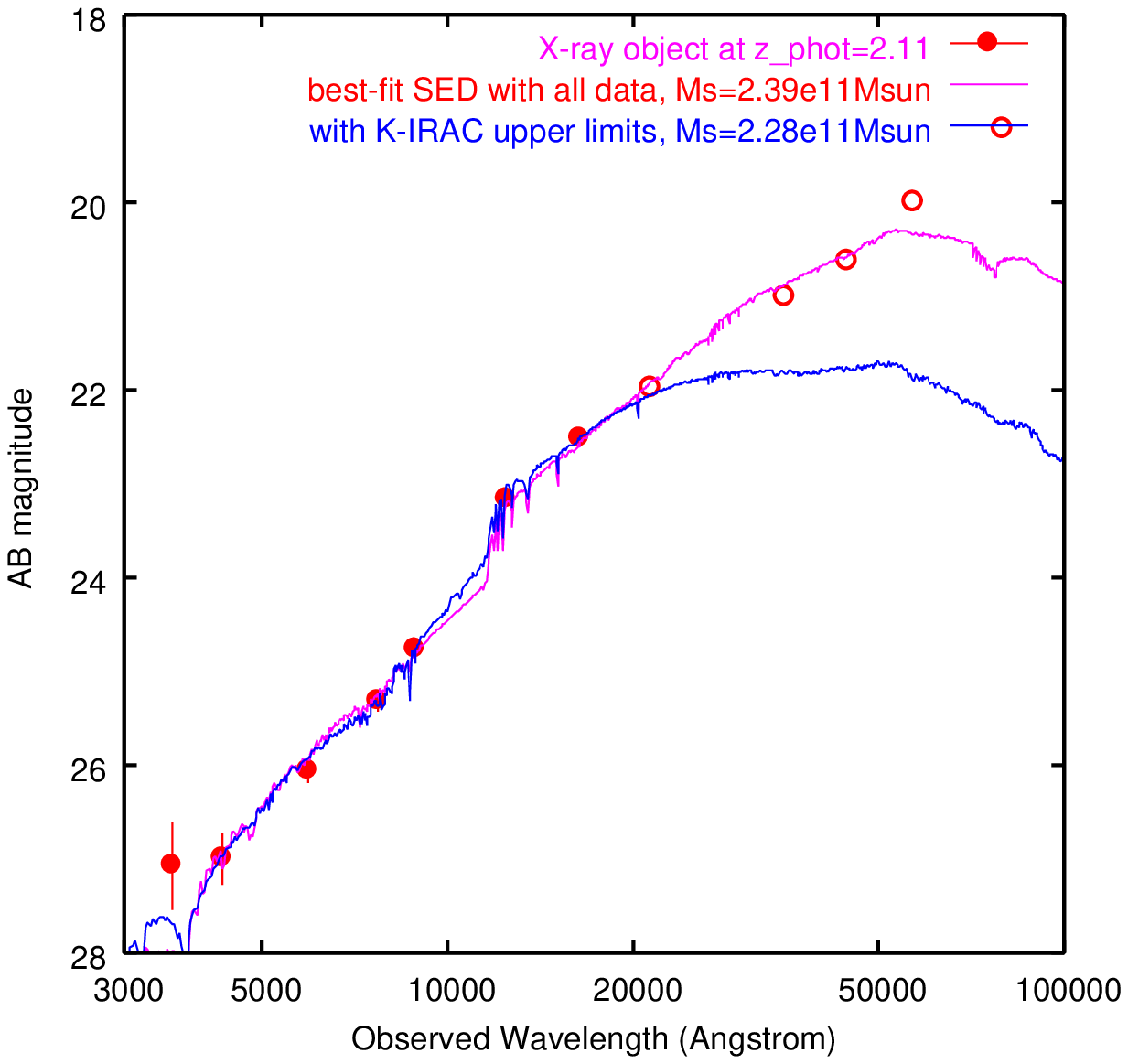}
\caption{ }
\label{fig4c}
\end{figure}

\figurenum{5}

\begin{figure}
\plotone{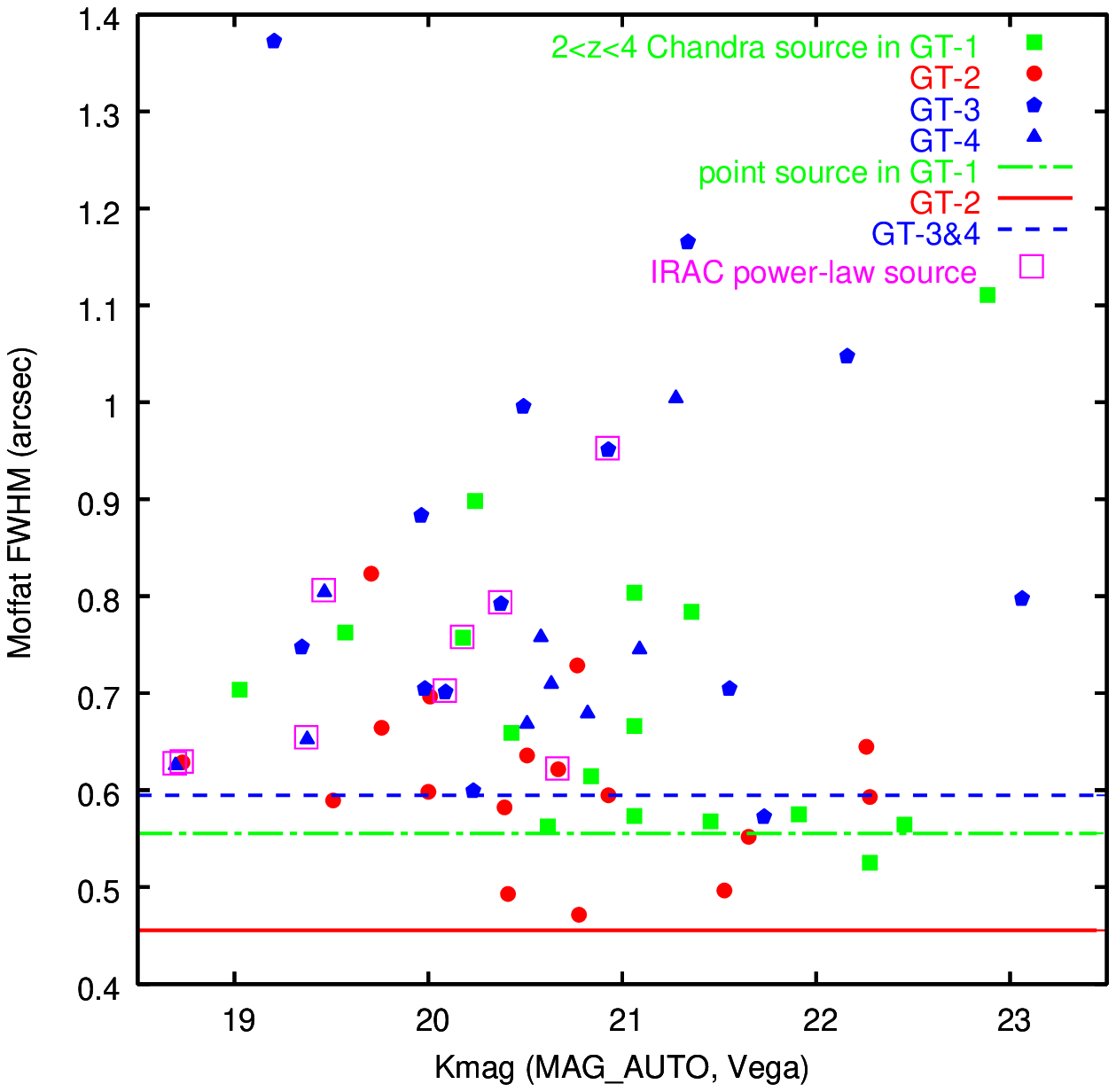}
\caption{The FWHM of the MOIRCS $Ks$-band images of the galaxies at $2 < z < 4$ which are detected in {\it Chandra} X-ray observations. Objects and the size of the stellar images in the different field of MODS are shown in the different color. The sources with 'power-law' SED (see text) are marked with the squares.}
\label{fig5}
\end{figure}
%
\figurenum{6}

\begin{figure}
\plotone{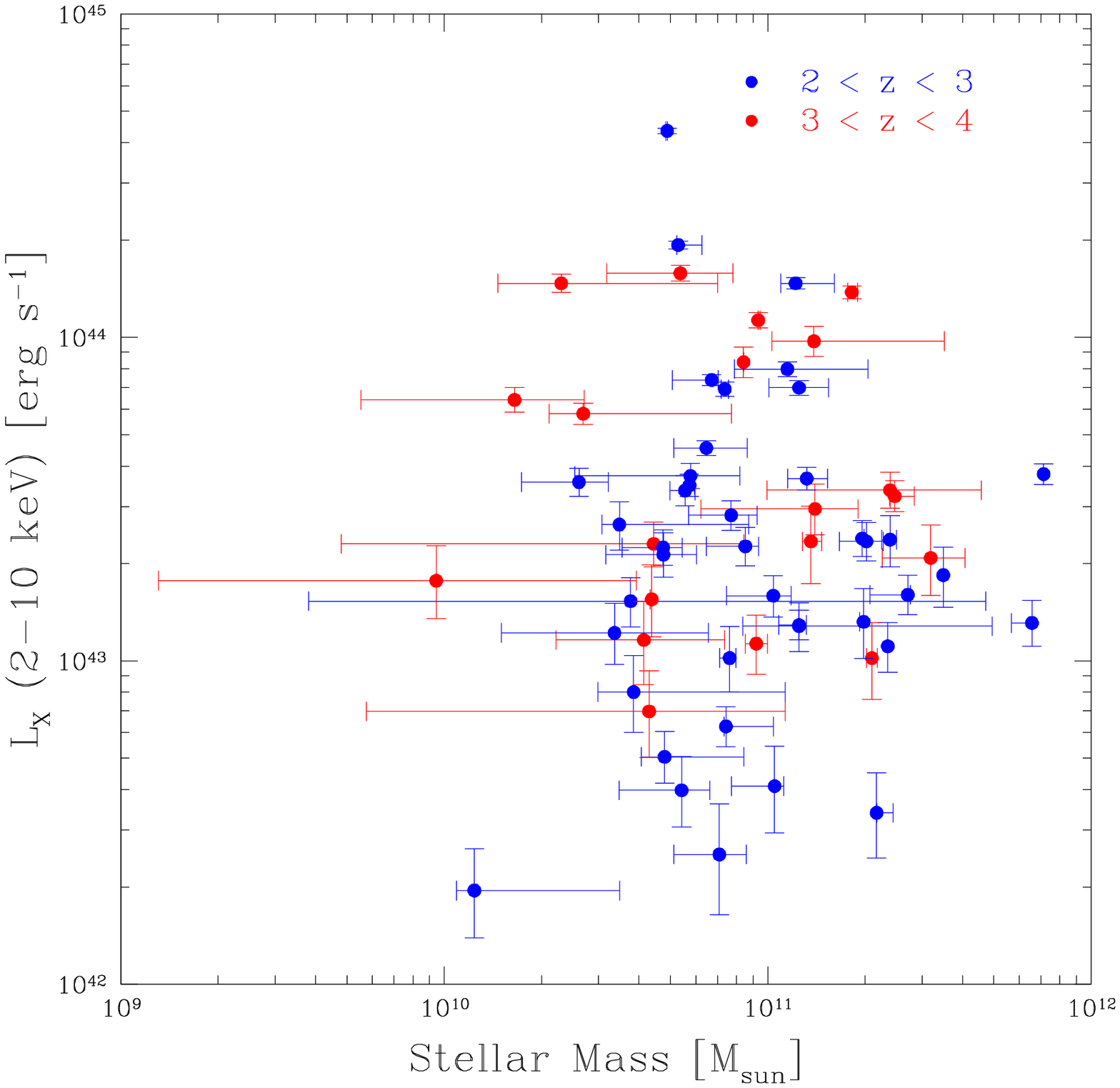}
\caption{X-ray luminosity at 2-10 keV as a function of the stellar mass of the $Ks$-band counterpart. The blue dots show the objects at $2 < z < 3$ and the red ones at $3 < z < 4$}
\label{fig6}
\end{figure}

\figurenum{7}

\begin{figure}
\plotone{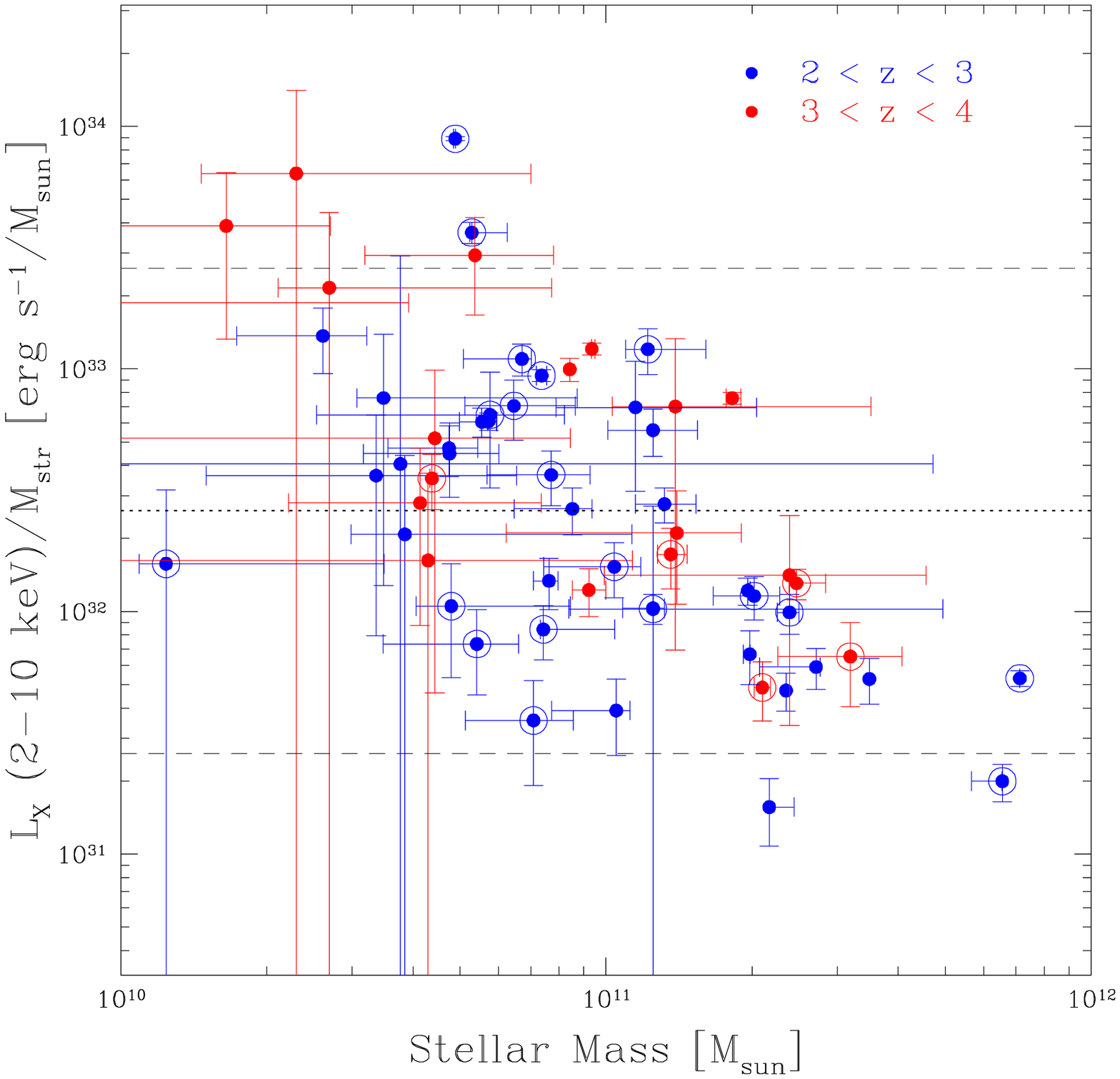}
\caption{Specific X-ray activity, namely the X-ray luminosity divided by the host stellar mass as a function of the stellar mass. The horizontal dotted line indicates the specific AGN activity in the case of M$_{\rm BH}$/M$_{\rm str}$=0.002, $\epsilon_X$=0.01, and $R_{\rm edd}=0.1$. The upper and lower dashed lines are for $R_{\rm edd}=1$ and 0.01, respectively. Large circles indicate the MIPS 24$\mu$m-band detection. }
\label{fig7}
\end{figure}

\figurenum{8}

\begin{figure}
\plotone{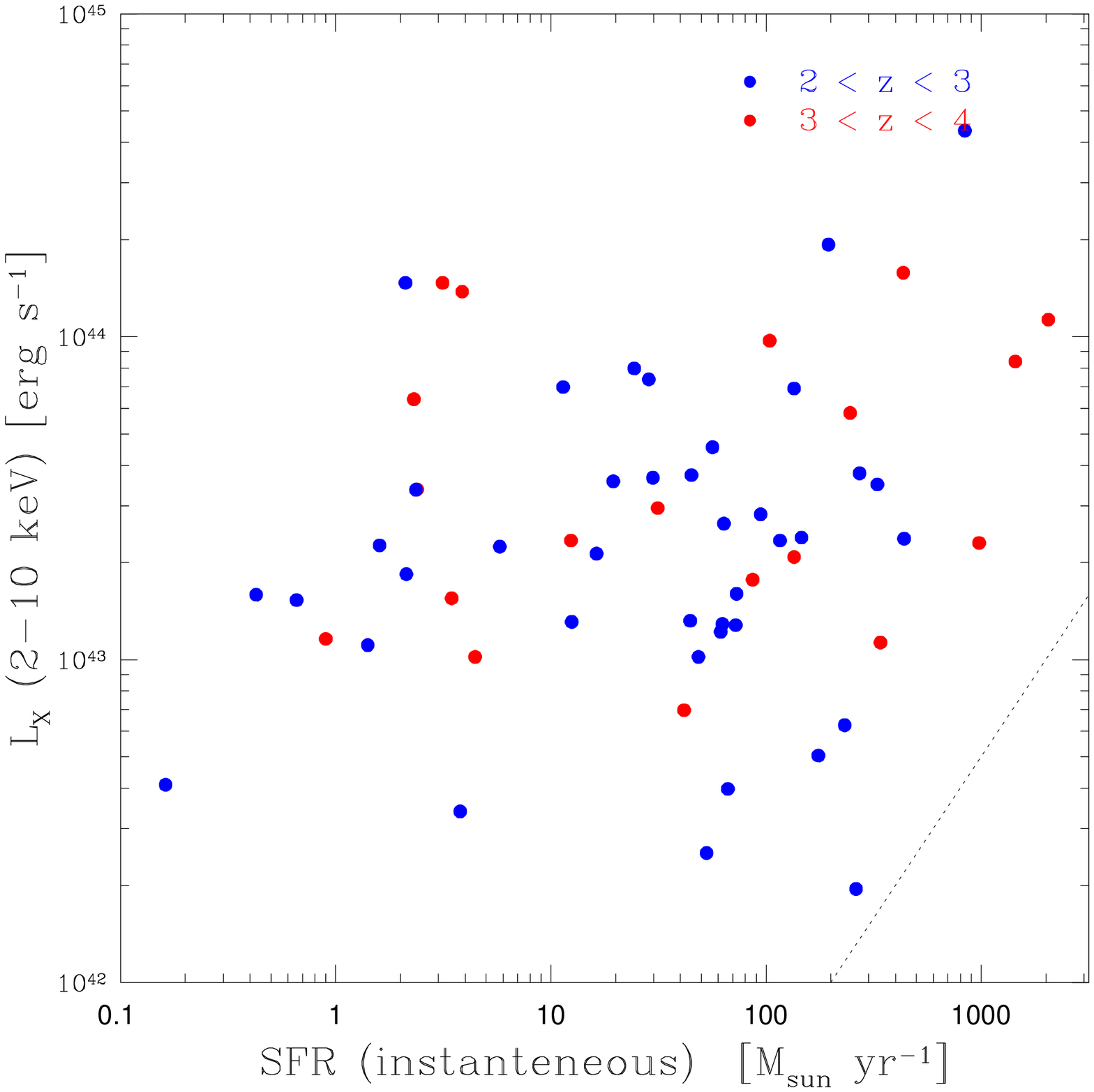}
\caption{X-ray luminosity as a function of the star-formation rate obtained in the SED fitting by the models. The instantaneous SFR is used in the left panel (a) while the SFR averaged over the past 300 Myr are used in the right panel (b). The dotted line is the expected X-ray luminosity from the star-formation component from the empirical fit by Ranalli {\it et al.} (2003).}
\label{fig8}
\end{figure}
%
\figurenum{8}
%
\begin{figure}
\plotone{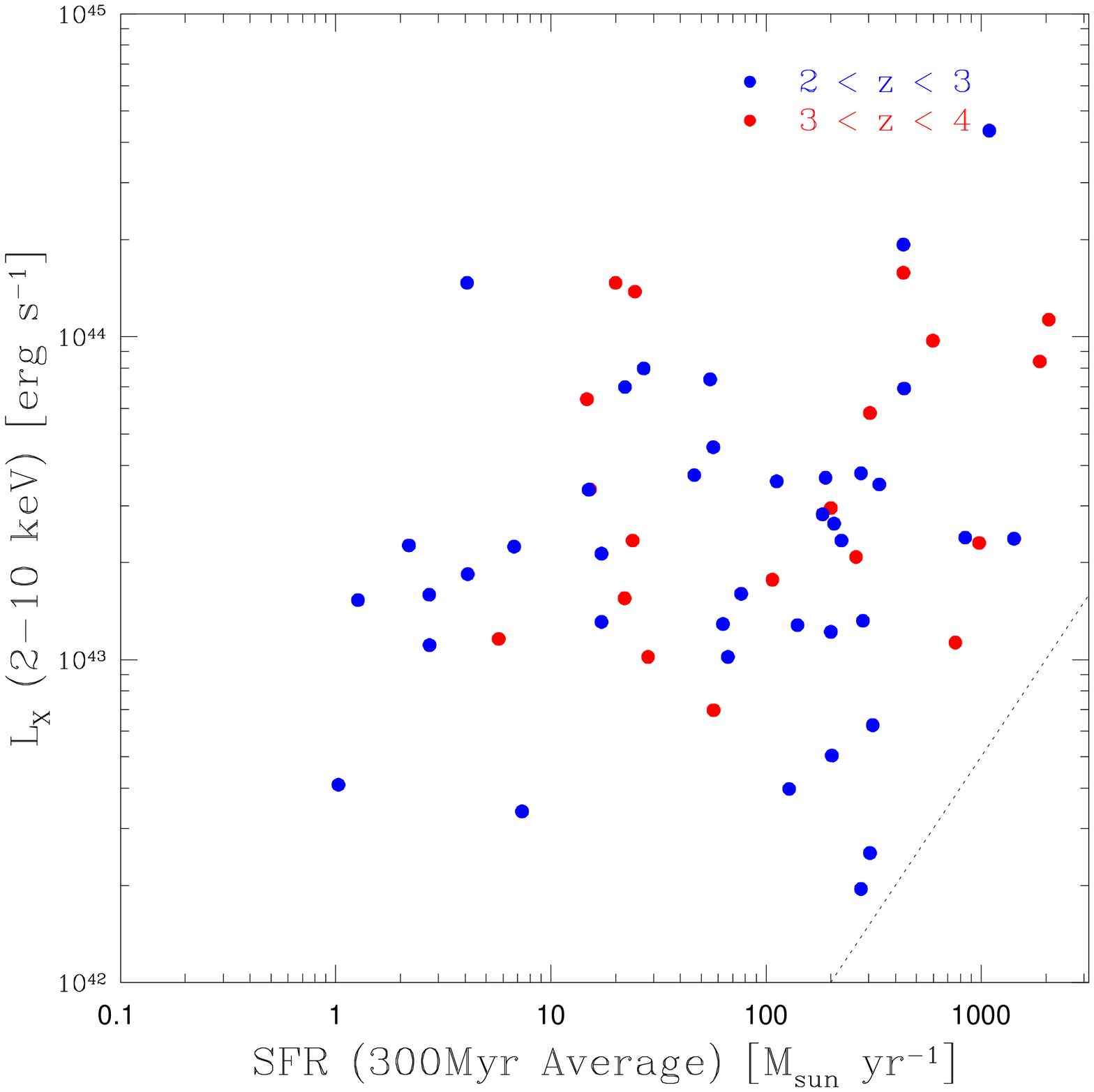}
\caption{(b)}
\label{fig8}
\end{figure}
%
\figurenum{9}

%
\begin{figure}
\plotone{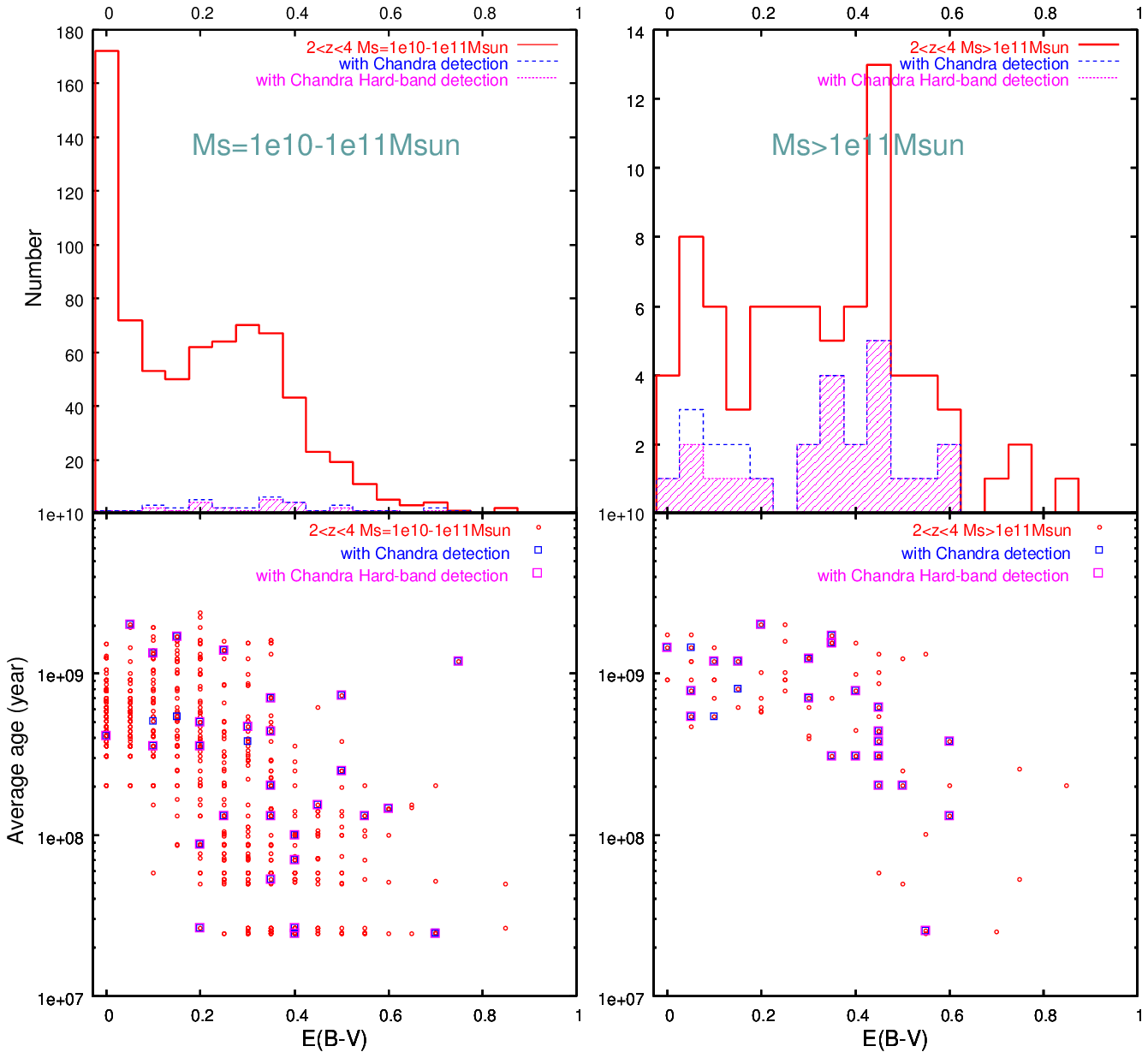}
\caption{The upper panels show the number distribution of the mass-selected galaxies with and without X-ray detection along the color excess values in the Calzetti extinction model which are obtained in the SED fitting. The lower panel show the average ages (see text) of the sample galaxies. Left and right panels show the sample with $10^{10}$ M$_\odot$ $<$ M$_{\rm str}$ $< 10^{11}$ M$_\odot$ and M$_{\rm str}$ $> 10^{11}$ M$_\odot$, respectively.
}
\label{fig9}
\end{figure}
%
%
\clearpage

\figurenum{10}

\begin{figure}
\plotone{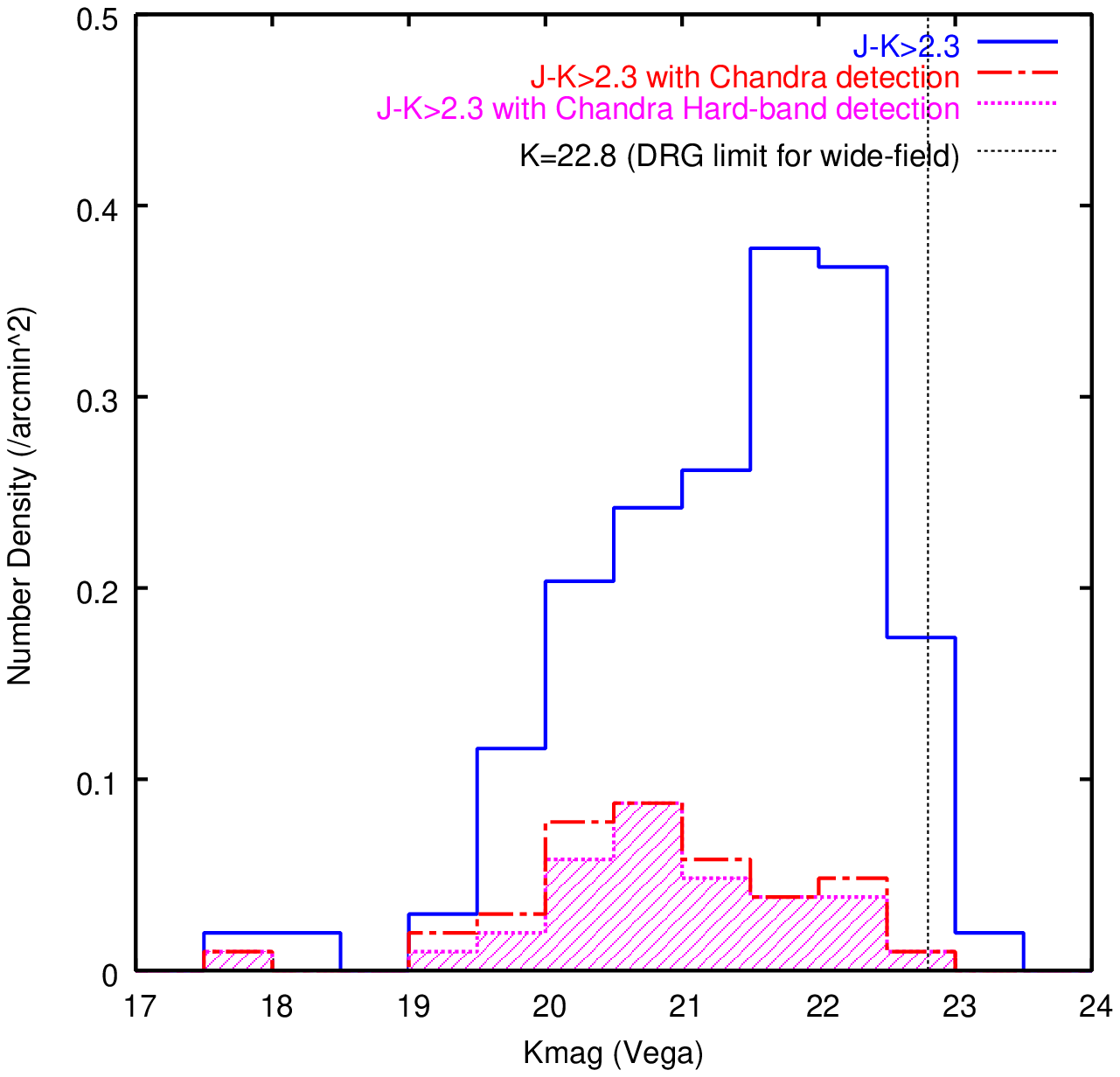}
\caption{The $K$-band number counts of all the DRGs as well as those identified with the X-ray sources. The counts of the objects associated with X-ray sources are shown by the lower histograms.
}
\label{fig10}
\end{figure}
%
\figurenum{11}
%
\begin{figure}
\plotone{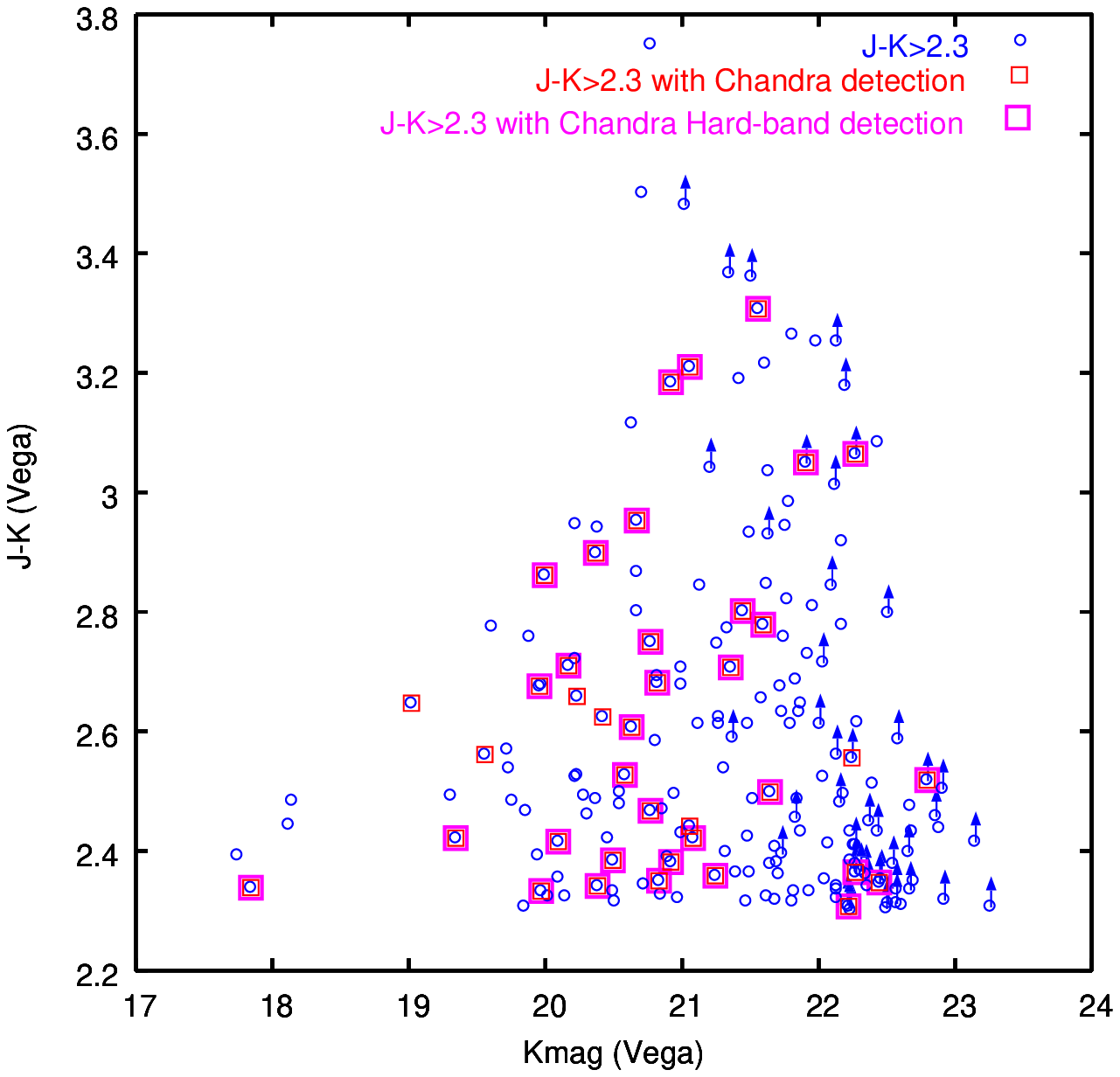}
\caption{The color-magnitude diagram of the DRGs in the MODS field. The objects detected in the hard-band and total-band are shown by the large and small squares, respectively.}
\label{fig11}
\end{figure}
%

\figurenum{12}
%
%
\begin{figure}
\plotone{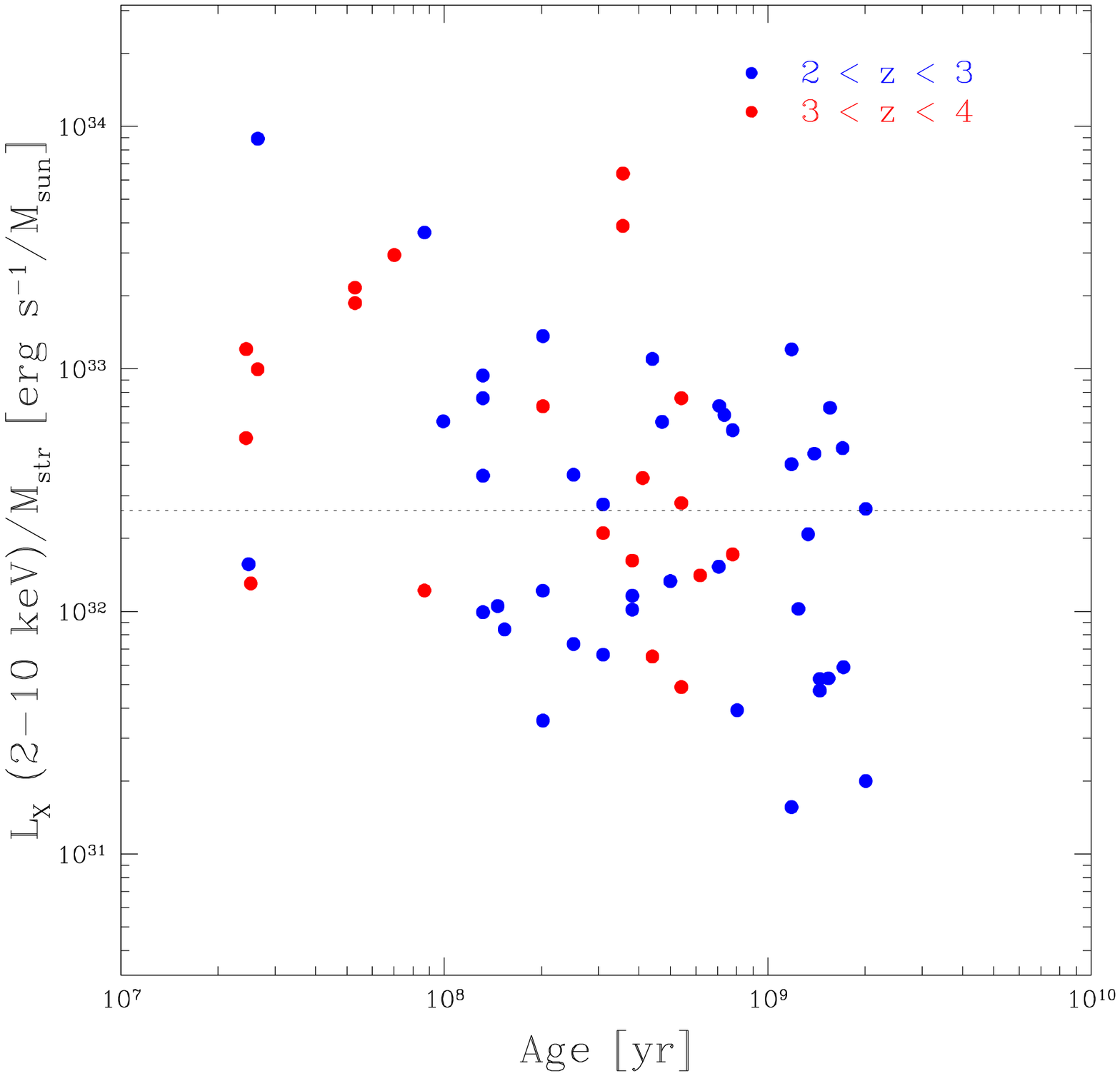}
\caption{The specific AGN activity versus the mean stellar ages obtained in the SED fitting. The blue dots show the objects at $2 < z < 3$ and the red ones at $3 < z < 4$. The horizontal dotted line is the same as in Fig.5.
}
\label{fig12}
\end{figure}

\end{document}